\documentclass[aps,preprint]{revtex4}%
\usepackage{amsfonts}
\usepackage{amsmath}
\usepackage{amssymb}
\usepackage{graphicx}%
\newtheorem{theorem}{Theorem}
\newtheorem{acknowledgement}[theorem]{Acknowledgement}

\begin{document}
\title{Far-reaching statistical consequences of the zero-point energy for the
harmonic oscillator}
\author{Luis de la Pe\~{n}a}
\email{luis@fisica.unam.mx}
\affiliation{Instituto de F\'{\i}sica, Universidad Nacional Aut\'{o}noma de M\'{e}xico
Apartado postal 20-364, 01000 M\'{e}xico}
\author{Andrea Vald\'{e}s-Hern\'{a}ndez}
\affiliation{Instituto de F\'{\i}sica, Universidad Nacional Aut\'{o}noma de M\'{e}xico
Apartado postal 20-364, 01000 M\'{e}xico}
\author{Ana Mar\'{\i}a Cetto}
\affiliation{Instituto de F\'{\i}sica, Universidad Nacional Aut\'{o}noma de M\'{e}xico
Apartado postal 20-364, 01000 M\'{e}xico\footnote{On temporary leave of
absence at the International Atomic Energy Agency, PO Box 200, 1200 Vienna,
Austria.}}

\begin{abstract}
In a recent thermodynamic analysis of the harmonic oscillator and using an
interpolation procedure, Boyer has shown that the existence of a zero-point
energy leads to the Planck spectrum. Here we avoid the interpolation by adding
a statistical argument to arrive at Planck's law as an inescapable result of
the presence of the zero-point energy. No explicit quantum argument is
introduced along the derivations. We disclose the connection of our results
with the original analysis of Planck and Einstein, which led to the notion of
the quantized radiation field. We then inquire into the discrete or continuous
behaviour of the energy and pinpoint the discontinuities. Finally, to open the
door to the description of the zero-point fluctuations, we briefly discuss the
statistical (in contrast to the purely thermodynamic) description of the
oscillator, which accounts for both thermal and temperature-independent
contributions to the energy dispersion.

\end{abstract}
\maketitle

\section{Introduction}

In a recent paper$^{\text{\cite{Boyer03}}}$ Boyer studies anew the
thermodynamics of the harmonic oscillator. By an elementary analysis based on
the invariance of the action under a quasistatic change of the frequency,
Boyer first reproduces Wien's displacement law (Eq. (\ref{12}) below). He then
introduces an important departure from usual treatments, by allowing for a
temperature-independent energy different from zero (the so-called zero-point
energy) in solving for the thermodynamic potential at low temperatures. This
constitutes an extension of the classical treatment that accords, however,
with quantum knowledge. The most immediate evidence of it is that it implies a
mean energy proportional to the oscillator's frequency at low temperatures, a
result that violates the equipartition principle but agrees with quantum theory.

To derive the Planck equilibrium law from the thermodynamic relations, Boyer
makes the smoothest possible interpolation between energy equipartition at
high temperatures and zero-point energy at low temperatures. In the present
work we show that this interpolation procedure can be avoided by adding a
statistical argument to derive the equilibrium spectrum from the sole
existence of the zero-point energy. Our approach leads unambiguosly to
Planck's law and reveals the decisive role of the zero-point energy in
defining the quantum behaviour of the system at equilibrium.

As is the case in Boyer's calculation, no explicit quantum argument is
introduced along the present derivation. After obtained our results we review
their connection with the original analysis of Planck$^{\text{\cite{Planck00}%
}}$ and Einstein,$^{\text{\cite{Einst09}}}$ which led to the notion of the
quantized radiation field (with no knowledge of the zero-point energy of
course), to show explicitly how much simplification and transparency is gained
by introducing the idea of the zero-point energy. We further inquire into the
discrete or continuous behavior of the energy; the analysis discloses the
origin and meaning of the discontinuities.

The thermodynamic analysis carried up to this point is limited, in that the
zero-point energy has a sure, nonfluctuating value. Therefore to conclude we
briefly discuss the statistical (in contrast to the purely thermodynamic)
description, which correctly accounts for both thermal and
temperature-independent contributions to the dispersion of the energy and
opens the door to the zero-point fluctuations.

\section{Basic relations}

\subsection{Thermodynamics of the harmonic oscillator}

In his thermodynamic analysis Boyer shows that the harmonic oscillator can be
described by means of a thermodynamic potential $\phi(z)$ (in Boyer's paper
the Boltzmann constant, $k$, is taken as 1 for simplicity)
\begin{equation}
\phi(z)=-\frac{1}{kT}F(\omega,T),\quad z=\frac{\omega}{T}, \label{10}%
\end{equation}
where $F(\omega,T)$ is the Helmholtz free energy. In particular, the average
oscillator energy in thermal equilibrium is given by%
\begin{equation}
U(\omega,T)=-\omega k\frac{d\phi(z)}{dz}=-\omega k\phi^{\prime}(z)=\omega
f(\omega/T), \label{12}%
\end{equation}
where the last equality corresponds to Wien's displacement law. The (thermal)
entropy of the oscillator is%
\begin{equation}
S(z)=k\phi(z)+\frac{1}{T}U(\omega,T), \label{14}%
\end{equation}
and for the specific heat at constant volume (and constant $\omega$) Boyer
obtains
\begin{equation}
C_{V}(z)=\left(  \frac{\partial U}{\partial T}\right)  _{V,\omega}=kz^{2}%
\phi^{\prime\prime}(z). \label{16}%
\end{equation}
These results are sufficient for our present purposes. However, before
proceeding let us use them to show how it is that the zero-point energy has
entered into the picture. In the low-temperature limit $(T\longrightarrow0)$,
Eq. (\ref{12}) reads%
\begin{equation}
U(\omega,0)\equiv\mathcal{E}_{0}=-\omega k\phi^{\prime}(\infty)=\text{const}%
\times\omega. \label{18}%
\end{equation}
In the usual thermodynamic analysis one \textit{arbitrarily} selects the
constant $\phi^{\prime}(\infty)=0$. Taking it as different from zero, which is
the more general possibility, we see the emergence of a zero-point energy that
is proportional to the frequency of the oscillator.$^{\text{\cite{relat}%
,\cite{unity}}}$ This selection, which runs contrary to energy equipartition
among the oscillators and hence to \textsc{19th} Century classical physics,
opens up interesting possibilities that we will explore, following Boyer.

\subsection{Thermodynamic distribution}

Our aim is to find the mean energy of the oscillators as a function of the
temperature, $\overline{E}=U(T).$ For this purpose we consider a system in
equilibrium at temperature $T$, and look for a distribution $W(E)$ subject to
the demand that the entropy be a maximum. Such entropy is defined through the
probability density $W$ by means of the relation
\begin{equation}
S=-k\int W\ln WdE. \label{1a}%
\end{equation}
The maximum entropy formalism$^{\text{\cite{mef}}}$ allows us to write the
general form of the probability that the energy acquires a value between $E$
and $E+dE$ as
\begin{subequations}
\label{22}%
\begin{align}
W_{g}(E)dE  &  =\frac{1}{Z_{g}(\beta)}g(E)e^{-\beta E}dE,\label{22a}\\
Z_{g}(\beta)  &  =\int g(E)e^{-\beta E}dE. \label{22b}%
\end{align}
Here $\beta$ is the inverse temperature, $\beta=1/(kT)$; $Z_{g}(\beta)$ is the
partition function and the factor $g(E)$ is the intrinsic probability of the
states with energy $E$. The distribution (\ref{22a}) is not new; already
Einstein considered it in his early works on the investigation of the specific
heat of solids.$^{\text{\cite{Einstein07}, \cite{note1}}}$

That this distribution is consistent with the thermodynamics derived by Boyer
can be easily seen by mere substitution of (\ref{22a}) in (\ref{1a}),
\end{subequations}
\begin{equation}
S=-k\int W_{g}(-\ln Z_{g}+\ln g-\beta E)dE=k\ln Z_{g}-k\overline{\ln g}+U/T.
\label{26}%
\end{equation}
Comparison with Eq. (\ref{14}) gives for the thermodynamic potential%
\begin{equation}
\phi=\ln Z_{g}-\overline{\ln g}. \label{27}%
\end{equation}
As shown in Appendix B the term $\overline{\ln g}$ is a numerical constant, so
that Eq. (\ref{27}) leads to%
\begin{equation}
\phi^{\prime}=\frac{d\phi}{dz}=\frac{1}{\omega k}\frac{1}{Z_{g}}\frac{\partial
Z_{g}}{\partial\beta}=-\frac{1}{\omega k}U, \label{32}%
\end{equation}
in agreement with Eq. (\ref{12}).

In standard classical theory all energies are assumed to have equal intrinsic
probabilities and thus $g(E)=1,$ which leads to
\begin{equation}
W_{g=1}(E)=\frac{1}{Z_{1}(\beta)}e^{-\beta E},\quad Z_{1}(\beta)=\int
_{0}^{\infty}e^{-\beta E}dE=\frac{1}{\beta},\quad\overline{E}=-\frac{1}{Z_{1}%
}\frac{dZ_{1}}{d\beta}=\frac{1}{\beta}, \label{23}%
\end{equation}
which is contrary to the existence of a zero-point energy. Consequently, in
order to allow for a zero-point energy different from zero we must resort to
the more general probability density (\ref{22a}), with $g(E)$ a function to be determined.

\section{Functional form for the mean energy}

\subsection{Basic statistical relations}

From Eqs. (\ref{22}) it follows that (from now on the prime indicates
derivative with respect to $\beta$)%
\begin{equation}
\overline{E^{r}}^{\prime}=-\frac{Z_{g}^{\prime}}{Z_{g}}\overline{E^{r}}%
-\frac{1}{Z}\int_{0}^{\infty}E^{r+1}g(E)e^{-\beta E}dE=-\frac{Z_{g}^{\prime}%
}{Z_{g}}\overline{E^{r}}-\overline{E^{r+1}}. \label{33}%
\end{equation}
Since
\begin{equation}
\overline{E}=\frac{1}{Z_{g}}\int_{0}^{\infty}Eg(E)e^{-\beta E}dE=-\frac
{Z_{g}^{\prime}}{Z_{g}}, \label{34}%
\end{equation}
Eq. (\ref{33}) gives the recurrence relation%
\begin{equation}
\overline{E^{r+1}}=\overline{E}\,\overline{E^{r}}-\overline{E^{r}}^{\prime}.
\label{36}%
\end{equation}
Incidentally this result can be extended to any continuous function $G(E)$, to
give
\begin{equation}
-\overline{G(E)}^{\prime}=\overline{EG(E)}-\overline{E\,}~\overline{\,G(E)},
\label{36b}%
\end{equation}
which shows that $-\overline{G(E)}^{\prime}$ is given\ in general by the
covariance of $G(E)$ and $E.\ $

In particular, for the second moment ($r=1$ in Eq. (\ref{36})) one gets for
the energy variance
\begin{equation}
\sigma_{E}^{2}\equiv\overline{E^{2}}-U^{2}=-U^{\prime}\text{.} \label{37}%
\end{equation}
This equation can be cast as the well known relation$^{\text{\cite{Huang}}}$
\begin{equation}
\sigma_{E}^{2}=-U^{\prime}=kT^{2}\left(  \frac{\partial U}{\partial T}\right)
_{V,\omega}=kT^{2}C_{V} \label{38}%
\end{equation}
in terms of the heat capacity at constant volume $C_{V}$. Since $C_{V}$
remains finite, the right hand side takes the value $0$ at $T=0$, whence%
\begin{equation}
\sigma_{E}^{2}(T=0)=0, \label{39}%
\end{equation}
which means that all fluctuations are supressed at zero temperature. This
result refers to \textit{thermal} fluctuations, since the description provided
by the distribution $W_{g}$ is of a thermodynamic nature ($W_{g}$ was
constructed by demanding consistency with the thermodynamic relations only).
The existence, description and origin of temperature-independent fluctuations
will be discussed below.

\subsection{Establishing the equilibrium spectrum\label{spectrum}}

Our task now is to determine the mean energy $U(\beta)$, using Eq. (\ref{37})
subject to the condition (\ref{39}). As follows from (\ref{37}), it is
possible to express $\sigma_{E}^{2}$ as a function of $U$ by inverting
$\overline{E(\beta)}=U$ to express $\beta$ as a function of the mean energy.
Assuming that the resulting expresion for $\sigma_{E}^{2}(U)$ admits a power
series expansion, we write it in the form%
\begin{equation}
\sigma_{E}^{2}(U)=%
{\textstyle\sum\limits_{n=0}}
a_{n}U^{n}, \label{370}%
\end{equation}
where the coefficients $a_{n}$ can depend only on the fixed parameter $\omega
$. Now we demand that the relative dispersion $\sigma_{E}/U$ remains finite
for all values of $U$, whence the expression%
\begin{equation}
\frac{\sigma_{E}^{2}}{U^{2}}=\frac{a_{0}}{U^{2}}+\frac{a_{1}}{U}+a_{2}%
+a_{3}U+a_{4}U^{2}+... \label{390}%
\end{equation}
must remain finite for every $U.$ In particular, since $U$ (being an
increasing function of $T$) can increase indefinitely, it follows that the
coefficients $a_{n}$ must vanish for $n\geq3,$ whence%
\begin{equation}
\sigma_{E}^{2}(U)=a_{0}+a_{1}U+a_{2}U^{2}. \label{400}%
\end{equation}
Further, if at low temperatures $U$ goes to zero, coefficient $a_{0}$ and
$a_{1}$ must be equal to zero for $\sigma_{E}/U$ to remain finite. However, if
at low temperatures $U$ does not vanish, the three coefficients can have in
principle a nonzero value.

We notice that the dispersion of the energy
\begin{equation}
\sigma_{E}^{2}=\left\langle E^{2}\right\rangle -\left\langle E\right\rangle
^{2} \label{525}%
\end{equation}
is invariant under the inversion $E\rightarrow-E.$ Since this transformation
induces the substitution $U\rightarrow-U,$ $\sigma_{E}^{2}$ must be an even
function of the variable $U,$ and therefore $a_{1}=0$ in (\ref{400}),%
\begin{equation}
\sigma_{E}^{2}(U)=a_{0}+a_{2}U^{2}. \label{400b}%
\end{equation}
In Appendix A we demonstrate that $a_{1}=0$ follows directly from Wien's law
for $\mathcal{E}_{0}>0.$ Combining this with Eq. (\ref{37}) we obtain%
\begin{equation}
\frac{dU}{a_{0}+a_{2}U^{2}}=-d\beta, \label{430}%
\end{equation}
which after integration gives%
\begin{equation}
\beta=\left\{
\begin{array}
[c]{ll}%
\frac{1}{a_{2}U} & \text{for }q=0;\\
\frac{2}{\sqrt{q}}\coth^{-1}\frac{2a_{2}U}{\sqrt{q}} & \text{for }q>0;
\end{array}
\right.  \qquad q\equiv-4a_{0}a_{2}. \label{440}%
\end{equation}
The case $q<0$ is excluded for real values of the energy (see equation
(\ref{410})\ below). Although the case $q=0$ can be treated as a limit case of
$q\geq0$, it is more illustrative to deal with the two cases
separately.\ Inverting the functions in (\ref{440}) we find%
\begin{equation}
U(\beta)=\left\{
\begin{array}
[c]{lc}%
\frac{1}{a_{2}\beta}, & \text{for }q=0;\\
\frac{\sqrt{q}}{2a_{2}}\coth\frac{\sqrt{q}}{2}\beta, & \text{for }q>0.
\end{array}
\right.  \label{445}%
\end{equation}
The behaviour of the mean energy is seen to depend critically on the value of
$q$, a parameter that appears naturally in the expression for the roots of the
equation $\sigma_{E}^{2}=0$,%
\begin{equation}
U_{\pm}=\pm\frac{\sqrt{q}}{2a_{2}}. \label{410}%
\end{equation}
Since $U(T=0)=\mathcal{E}_{0}$, it follows from Eq. (\ref{39}) that
$\sigma_{E}^{2}(\mathcal{E}_{0})=0,$ so $\mathcal{E}_{0}$ is certainly one of
the roots $U_{\pm}.$ Further, since the dispersion is an increasing function
of the energy ($U(\beta)\geq\mathcal{E}_{0}$) it follows that $a_{2}>0,$ and
so $\mathcal{E}_{0}$ is given by
\begin{equation}
\mathcal{E}_{0}=\frac{\sqrt{q}}{2a_{2}}=\sqrt{-\frac{a_{0}}{a_{2}}}
\label{420}%
\end{equation}
(the other root being unphysical). This relation between $q$ and
$\mathcal{E}_{0}$, along with Eq. (\ref{445}), shows how the functional form
of the mean energy is uniquely determined by the zero-point energy. In
particular, for a theory with null value for $\mathcal{E}_{0}$ we have $q=0$,
$a_{0}=0$ and from (\ref{445}), $U=(a_{2}\beta)^{-1}$. Comparison with the
classical case compels us to set $a_{2}=1,$ whence%
\begin{equation}
U=kT. \label{421}%
\end{equation}
\qquad\qquad If, however, the theory allows for a zero-point energy
$\mathcal{E}_{0}\neq0$, $q$ acquires a value different from $0$ as follows
from Eq. (\ref{420}). The latter equation, together with Eq. (\ref{445}),
leads to
\begin{equation}
U(\beta)=\mathcal{E}_{0}\coth a_{2}\mathcal{E}_{0}\beta. \label{508}%
\end{equation}
Further, by taking the limit $T\rightarrow\infty$ $(\beta\rightarrow0)$ we get%
\begin{equation}
U(\beta\rightarrow0)=\frac{1}{a_{2}\beta}. \label{510}%
\end{equation}
The condition that at high temperatures the result (\ref{508}) coincides with
the classical one fixes $a_{2}=1$ as before, whence we finally have%
\begin{equation}
U(\beta)=\mathcal{E}_{0}\coth\mathcal{E}_{0}\beta. \label{520}%
\end{equation}
This is Planck's law with zero-point energy included, as follows by taking the
zero-temperature limit $T\rightarrow0$ $(\beta\rightarrow\infty)$,%
\begin{equation}
U(\beta\rightarrow\infty)=\mathcal{E}_{0}. \label{521}%
\end{equation}
and recalling from Eq. (\ref{18})\ that $\mathcal{E}_{0}=$const$\times\omega$.

This establishes Planck's spectral distribution law as a physical result whose
ultimate meaning is the existence of a zero-point energy, whereas the
equipartition of energy reflects its absence, in full accordance with Boyer's
remarks.$^{\text{\cite{Boyer03}}}$

It is important to stress that Planck's law has been obtained without
introducing any \emph{explicit} quantum demand. However, the fact that the law
that gave birth to quantum theory stems from the existence of a zero-point
energy, brings to the fore the crucial importance of this
temperature-independent energy for the understanding of quantum mechanics.

\section{Planck, Einstein and the zero-point energy\label{discussion}}

Having fixed the parameter $a_{2}=1$, the value of $a_{0}$ follows from Eq.
(\ref{420})%
\begin{equation}
q=4\mathcal{E}_{0}^{2},\quad a_{0}=-\mathcal{E}_{0}^{2}. \label{526}%
\end{equation}
Substitution of these values in Eq. (\ref{400b}) gives for the energy dispersion%

\begin{equation}
\sigma_{E}^{2}(U)=U^{2}-\mathcal{E}_{0}{}^{2}\quad(U=U_{\text{Planck}}),
\label{521b}%
\end{equation}
whereas in the classical case $q=0,$ $\mathcal{E}_{0}=0$ and%

\begin{equation}
\sigma_{E}^{2}(U)=U^{2}\quad(U=U_{\text{equipartition}}). \label{521c}%
\end{equation}

Whilst in the latter case the thermal fluctuations of the oscillator's energy
depend on its thermal mean energy, in the quantum case they are expressed in
terms of the total mean energy including a temperature-independent
contribution, according to Eq. (\ref{521b}). At $T=0$ however, the thermal
fluctuations of the energy vanish in the quantum case just as in the classical
case --- which means that any fluctuation of the energy at zero temperature
must be non-thermal. Below we show that a result similar in form to Eq.
(\ref{521c}) holds good in the general case, once all (thermal and
non-thermal) fluctuations and energies are taken into account by using a
complete statistical description. Since $\sigma_{E}^{2}$ in Eq. (\ref{521b})
contains \emph{only} thermal fluctuations, it means that $\mathcal{E}_{0}%
{}^{2}$ stands for the fluctuations of the zero-point energy, a result that is
verified by Eq. (\ref{521c}).

The above discussion suggests the time-honoured convention of separating the
average energy $U$ into a thermal $U_{T}$ and a temperature-independent
$\mathcal{E}_{0}$ contribution,
\begin{equation}
U=U_{T}+\mathcal{E}_{0}. \label{540}%
\end{equation}
Substitution in (\ref{521b}) then gives%
\begin{equation}
\sigma_{E}^{2}=U_{T}^{2}+2\mathcal{E}_{0}U_{T}, \label{550}%
\end{equation}
and since $dU/d\beta=dU_{T}/d\beta$, it follows using Eq. (\ref{37}) that
\begin{equation}
-\frac{dU_{T}}{d\beta}=U_{T}^{2}+2\mathcal{E}_{0}U_{T}. \label{555}%
\end{equation}
No wonder that the appropiate solution to this equation for $\mathcal{E}%
_{0}\neq0$ is the Planck spectrum without the zero-point term,%
\begin{equation}
U_{T}=\frac{2\mathcal{E}_{0}}{e^{2\mathcal{E}_{0}\beta}-1}. \label{556}%
\end{equation}
At sufficiently low temperatures $(\beta\rightarrow\infty)$ this solution
takes the form%
\begin{equation}
U_{T}=2\mathcal{E}_{0}e^{-2\mathcal{E}_{0}\beta}, \label{560}%
\end{equation}
which is the (approximate) distribution suggested by Wien at the end of the
19\textsc{th} century.

Equations (\ref{550}) and (\ref{560}) were decisive for the initial
construction of quantum theory since they led Einstein and Planck to establish
the existence of quanta. Let us now briefly discuss the above results and the
role played by the then hidden zero-point energy, by paying attention to two
fundamental moments in the development of quantum
theory.$^{\text{\cite{PeCe02}}}$

\subsection{Planck's analysis...}

In his initial studies on the radiation field in equilibrium with matter,
Planck$^{\text{\cite{Planck00}}}$ used as point of departure the expression
\begin{equation}
\frac{\partial S}{\partial U}=\frac{1}{T}. \label{110}%
\end{equation}
In agreement with the then classical views that recognized only a thermal
energy, there was no room for a zero-point energy, so that $U$ should be
replaced by $U_{T}$. In the high-temperature limit the relation (\ref{110})
led Planck to write (taking $U_{T}\left(  T\rightarrow\infty\right)  =kT$)
\begin{equation}
\frac{\partial^{2}S}{\partial U_{T}^{2}}=\frac{\partial}{\partial U_{T}%
}\left(  \frac{k}{U_{T}}\right)  =-\frac{k}{U_{T}^{2}}. \label{112}%
\end{equation}
However, for his description of the low-temperature behaviour Planck used
Wien's law Eq. (\ref{560}), believed in those days to be an exact description
of the properties of the equilibrium field (at the low temperatures tested at
the time). He thus wrote (using modern notation, with $\mathcal{E}_{0}%
\ $=$\hbar\omega/2$)
\begin{equation}
U_{T}=2\mathcal{E}_{0}e^{-2\mathcal{E}_{0}\beta}=2\mathcal{E}_{0}%
e^{-2\mathcal{E}_{0}/kT}=2\mathcal{E}_{0}e^{-2\left(  \mathcal{E}%
_{0}/k\right)  \left(  \partial S/\partial U_{T}\right)  }, \label{114a}%
\end{equation}
whence%

\begin{subequations}
\begin{equation}
\frac{\partial S}{\partial U_{T}}=-\frac{k}{2\mathcal{E}_{0}}\ln\frac{U_{T}%
}{2\mathcal{E}_{0}} \label{114b}%
\end{equation}
and
\begin{equation}
\frac{\partial^{2}S}{\partial U_{T}^{2}}=-\frac{k}{2\mathcal{E}_{0}U_{T}}.
\label{114c}%
\end{equation}
As is well known, Planck correctly assumed that the description for arbitrary
temperature could be obtained from a direct interpolation of Eqs. (\ref{112})
and (\ref{114c}), so he proposed to write
\end{subequations}
\begin{equation}
\frac{\partial^{2}S}{\partial U_{T}^{2}}=-\frac{k}{U_{T}^{2}+2\mathcal{E}%
_{0}U_{T}}. \label{116}%
\end{equation}
This leads immediately to Planck's law without zero-point term (Eq.
(\ref{556})), a result that Planck interpreted afterwards as due to the
quantization of the interchanged energy between the material oscillators and
the equilibrium radiation field.

\subsection{Einstein's ouverture...}

A few years thereafter, Einstein argued that Eq. (\ref{116}) was well
confirmed by experiment and should therefore be used instead of previous
flawed alternatives, though its meaning remained to be clarified. From Eq.
(\ref{110}), which Einstein took as a secure point of departure stemming from
thermodynamics,
\begin{equation}
\frac{\partial^{2}S}{\partial U_{T}^{2}}=\frac{\partial}{\partial U_{T}}%
\frac{1}{T}=-\frac{1}{T^{2}C_{V}}, \label{118a}%
\end{equation}
one obtains%
\begin{equation}
kT^{2}C_{V}=-k\left(  \frac{\partial^{2}S}{\partial U_{T}^{2}}\right)  ^{-1}.
\label{118b}%
\end{equation}
Combining this with Eqs. (\ref{116}) and (\ref{38}) one is led to
\begin{equation}
kT^{2}C_{V}=-\frac{\partial U_{T}}{\partial\beta}=\sigma_{E}^{2}=U_{T}%
^{2}+2\mathcal{E}_{0}U_{T}, \label{120}%
\end{equation}
which reproduces Eq. (\ref{550}). Because this expression is at variance with
the classical result $\sigma_{E}^{2}=U_{T}^{2},$ Einstein took upon himself to
decipher its meaning. As is frequently stated, it is here where he brought off
his most (according to him, his only) revolutionary step in physics. He
interpreted the first term on the right hand side of Eq. (\ref{120}) as due to
the fluctuations of the thermal field (of a given frequency) produced by the
interferences among its modes. This interpretation follows from considering
the limit of Eq. (\ref{120}) for high temperatures, for which $U_{T}%
\gg\mathcal{E}_{0}$, and therefore $\sigma_{E}^{2}=U_{T}^{2},$ as predicted by
Maxwell's equations (without zero-point energy, of
course).$^{\text{\cite{vedral}}}$ Thus Einstein saw in this term a direct
manifestation of the undulatory nature of light.

As for the second term in (\ref{120}) --- unexpected from classical
thermodynamics --- the fact that it leads to the (quantum) theory of Planck
induced Einstein to interpret it in terms of light quanta; that is, he saw in
the expression $2\mathcal{E}_{0}U_{T}$ the manifestation of a discrete
property of the radiation field. According to Planck, the average interchanged
energy between the material oscillators and the radiation field is $\Delta
U=\hbar\omega\overline{n}$ and the extra fluctuations contribute with a
variance $\sigma_{\Delta U}^{2}=2\mathcal{E}_{0}\Delta U=\hbar^{2}\omega
^{2}\overline{n}$, as follows from Eq. (\ref{120}). Einstein argued that the
linear character of the variance in $\overline{n}$ suggests a Poisson
distribution that describes $n$ independent events, each one interchanging an
energy equal to $\hbar\omega.$ Thus Einstein interpreted the linear term as a
"corpuscular" contribution of the field, each corpuscle being an independent
packet of energy $\hbar\omega$ ---the photon, in our modern
parlance.$^{\text{\cite{vedral}}}$ This was indeed the birth of the photon
theory. It is clear from Eq. (\ref{120}) that such discrete structure of the
field will manifest itself at very low temperatures, where the linear term
dominates over the quadratic, undulatory one. However, as stressed by Einstein
from 1909 onwards, both terms coexist at all temperatures, and thus both
particle and wave properties of the light coexist.$^{\text{\cite{Einst09}}}$

\subsection{... and the zero-point energy}

At this stage it is interesting to make some comments on Einstein's analysis
of Eq. (\ref{120}). According to Einstein, the first term on the right hand
side $(U_{T}^{2})$ is a manifestation of the undulatory nature of the
(monochromatic) thermal field, whereas the second one $(2\mathcal{E}_{0}%
U_{T})$ reflects its quantum aspect. No zero-point energy is considered and a
corpuscular property of the radiation field emerges. However, we have seen
that the acceptance of a zero-point energy gives rise to an alternative
understanding of Eq. (\ref{120}). By allowing for a zero-point contribution to
the energy, the interference interpretation of the term $U_{T}^{2}$ suggests
to understand the term $2\mathcal{E}_{0}U_{T}$ as due to additional
interferences between the thermal field and a zero-point field ultimately
responsible for the zero-point energy. There is no extra contribution
$\mathcal{E}_{0}^{2}$ in Eq. (\ref{120}) that stands for the interferences
among the modes of the zero-point field itself because the present
thermodynamic description has no room for the temperature-independent
fluctuations of the zero-point energy, as has already been pointed out.

We see that within this restricted approach, no intrinsic discontinuities in
the field or in the interchange of energy are needed to derive Planck's law,
the existence of a zero-point energy being enough to understand the
equilibrium spectrum that follows from Eqs. (\ref{116}) and (\ref{120}). This,
of course, could not be Planck's or Einstein's interpretation since the
zero-point energy was still unknown at that time, and Planck deemed himself
forced to introduce the notion of quantization. Anyhow, we have here three
different perspectives on the same quantity $U_{T}^{2}+2\mathcal{E}_{0}U_{T}.$

As said above, Einstein was led for the first time$^{\text{\cite{Einst05}}}$
to his photon theory by the term $2\mathcal{E}_{0}U_{T},$ which he got from
Wien's law Eq. (\ref{560}). This approximate expression already contains the
constant $\mathcal{E}_{0}$, through which the seed of the zero-point energy
was planted in the results obtained by Planck and Einstein, although due to
the circumstances they were unable to interpret them in such terms. Our
discussion reinforces the conclusion reached in the previous section about the
univocal relation between the zero-point energy and the Planck equilibrium
spectrum, bolstering at the same time the need to inquire about the
relationship between the zero-point energy and quantization. This we take as
our next subject.

\section{Continuous \textit{vs} discrete}

As seen from Eqs. (\ref{116}) and (\ref{120}), both Planck and Einstein came
up with the binomial $U_{T}^{2}+2\mathcal{E}_{0}U_{T}$ whose structure was
eventually attributed to a discrete, corpuscle-like property of the radiation
field. Within the present approach, by contrast, this expression can be
interpreted as a direct result of the existence of the zero-point energy, with
no apparent need to resort to discontinuities. Our discussion thus suggests
that Planck and Einstein were unknowingly using the zero-point energy,
concealed by the classical thermodynamic treatment and accounted for by the
quantum description. This is a most inspiring hallmark, which leads us to
analyse the apparent contradiction between both approaches: Is quantization
necessarily linked to Planck's law, or is it not? Is it merely the result of a
point of view, or does it reflect and describe a true, ontological property of nature?

\subsection{The partition function}

As follows from Eq. (\ref{34}), once $U(\beta)$ is known it is possible to
determine the partition function $Z_{g}(\beta)$ by direct integration of%
\begin{equation}
U=-\frac{d\ln Z_{g}(\beta)}{d\beta}. \label{710}%
\end{equation}
Substituting Eq. (\ref{520}) gives%
\[
\int\mathcal{E}_{0}\coth\mathcal{E}_{0}\beta~d\beta=-\ln Z_{g}+\ln C,
\]
which leads to%
\begin{equation}
Z_{g}=\frac{C}{\sinh\mathcal{E}_{0}\beta}. \label{78}%
\end{equation}
The value of the constant $C$ can be determined by demanding that in the limit
$T\rightarrow\infty$ the classical result $Z_{g}(\beta\rightarrow0)=\beta
^{-1}$ be recovered. This leads to $C=\mathcal{E}_{0}$ and
\begin{equation}
Z_{g}(\beta)=\frac{\mathcal{E}_{0}}{\sinh\mathcal{E}_{0}\beta}. \label{79}%
\end{equation}
As follows from Eq. (\ref{26}), the entropy of the system is given by
($z=\omega/T$)
\begin{equation}
S(z)=k\phi(z)+k\beta U(\omega,T)+c=-k\ln\left(  \sinh\mathcal{E}_{0}%
\beta\right)  +k\mathcal{E}_{0}\beta\coth\mathcal{E}_{0}\beta+k\ln
\mathcal{E}_{0}-k\overline{\ln g}+c, \label{81}%
\end{equation}
where the value of the additive constant $c$ is determined by setting
$S(T=0)=0,$
\begin{equation}
c=-k\ln2\mathcal{E}_{0}+k\overline{\ln g}, \label{82}%
\end{equation}
and therefore Eq. (\ref{81}) reduces to%
\begin{equation}
S(z)=-k\ln\left(  2\sinh\mathcal{E}_{0}\beta\right)  +k\mathcal{E}_{0}%
\beta\coth\mathcal{E}_{0}\beta, \label{83}%
\end{equation}
which coincides with the result reported in Boyer's paper (with $k=1$%
).$^{\text{\cite{Boyer03}}}$

\subsection{The origin of discreteness\emph{\ }}

Let us now proceed to reveal the discontinuities characteristic of the quantum
description, which are hidden under the fully continuous description afforded
by the distribution $W_{g}$. To this aim we expand Eq. (\ref{79}) and
write$^{\text{\cite{STL}}}$%
\begin{equation}
Z_{g}=2\mathcal{E}_{0}\frac{1}{2\sinh\mathcal{E}_{0}\beta}=2\mathcal{E}%
_{0}\frac{e^{-\mathcal{E}_{0}\beta}}{1-e^{-2\mathcal{E}_{0}\beta}%
}=2\mathcal{E}_{0}\sum_{n=0}^{\infty}e^{-\mathcal{E}_{0}\beta(2n+1)},
\label{130}%
\end{equation}
or%
\begin{equation}
Z_{g}=2\mathcal{E}_{0}\sum_{n=0}^{\infty}e^{-\beta E_{n}},\quad E_{n}%
\equiv(2n+1)\mathcal{E}_{0}. \label{132}%
\end{equation}
This expression allows us to determine the function $g(E)$ by means of the
relation (\ref{22b}),
\begin{subequations}
\begin{equation}
Z_{g}(\beta)=\int_{0}^{\infty}g(E)e^{-\beta E}dE=2\mathcal{E}_{0}\sum
_{n=0}^{\infty}e^{-\beta E_{n}}\equiv2\mathcal{E}_{0}Z, \label{136a}%
\end{equation}
where we have introduced the dimensionless partition function
\begin{equation}
Z=\sum_{n=0}^{\infty}e^{-\beta E_{n}}=\frac{1}{2\sinh\mathcal{E}_{0}\beta}.
\label{136aa}%
\end{equation}
Thus $g(E)$\ can be cast as
\end{subequations}
\begin{equation}
g(E)=2\mathcal{E}_{0}\sum_{n=0}^{\infty}\delta(E-E_{n}). \label{136b}%
\end{equation}
Substitution of (\ref{136a}) and (\ref{136b}) in (\ref{22a}) results in%
\begin{equation}
W_{g}(E)=\frac{1}{Z}\sum_{n=0}^{\infty}\delta(E-E_{n})e^{-\beta E}.
\label{138a}%
\end{equation}
This distribution gives for the mean value of any function $f(E)$
\begin{equation}
\overline{f(E)}=\int_{0}^{\infty}W_{g}(E)f(E)dE=\frac{1}{Z}\sum_{n=0}^{\infty
}f(E_{n})e^{-\beta E_{n}}=\sum_{n=0}^{\infty}w_{n}f(E_{n}), \label{138c}%
\end{equation}
where we have introduced the weights (relative probabilities)
\begin{equation}
w_{n}=\frac{e^{-\beta E_{n}}}{Z}=\frac{e^{-\beta E_{n}}}{\sum_{n=0}^{\infty
}e^{-\beta E_{n}}}. \label{139}%
\end{equation}
Eq. (\ref{138c}) shows that the mean value of any function of the continuous
variable $E$ weighted with the distribution $W_{g}(E),$ can equivalently be
written as an average weighted with $w_{n}$ over a set of discrete indices
$n$. Since we are describing a canonical ensemble, the structure of $w_{n}$
suggests to identify the quantity $E_{n}$ with discrete energy levels of the
quantum oscillators including of course the zero-point energy, as follows from
(\ref{132}). Thus we can recognize in Eq. (\ref{139}) the description afforded
by the density matrix for the canonical ensemble with weights $w_{n}%
$.$^{\text{\cite{CTDL}}}$

Even though both averages (those calculated by means of $W_{g}$ and $w_{n}$)
are formally equivalent, it must be pointed out that the descriptions afforded
by each of these distributions are essentially different, referring to a
continuous or a discrete energy, respectively. The expansion in Eq.
(\ref{138c}) allows to pass from a description involving an averaging over the
in principle continuous variable $E$ to another one involving a summation over
discrete \emph{states} $n$. Since the energy in this new context $(E_{n})$ is
completely characterized by these states, it becomes natural to interpret the
right hand side of (\ref{138c}) as a manifestation of the discrete nature of
the energy. The mechanism leading to them, seemingly excluding all other
values of the energy, is of course identified with the highly patological
distribution $g(E)$.

These observations show how deeply the introduction of a zero-point energy
agrees with the quantum notion introduced by Planck and Einstein, and serves
to discover the fundamental role of the zero-point energy in explaining
quantization, it being at the root of Eq. (\ref{130}) and hence of (\ref{136b}).

\section{A quantum statistical distribution}

The analysis just presented leads us to conclude that although $E$ is a
continuous variable, its mean values corresponding to the thermodynamic
equilibrium states of a canonical ensemble of oscillators are extremely peaked
and approximate very closely a discrete spectrum. That is to say, the energies
that conform the thermal equilibrium state described by the distribution
$W_{g}$ belong, roughly speaking, to a discrete spectrum. This explains why
the mean value $\overline{f(E)}$ --- that corresponds to an equilibrium state
--- involves only the discrete set $E_{n}$.

To this, however, we should add that nevertheless the energy fluctuates, and
can therefore acquire values from among a continuous
spectrum.$^{\text{\cite{note2}}}$ As was stated at the beginning of Section
\ref{discussion}, the existence of a zero-point energy in the thermodynamic
description of the harmonic oscillator demands looking for a more general
distribution (instead of $W_{g})$ that could account for \emph{all}
fluctuations of the energy, including any temperature-independent
contribution. That such distribution must exist follows from the previous
results showing that the zero-point energy approach (starting from a
continuous energy distribution) led to a result that is equivalent to the
quantum description, in which temperature-independent fluctuations appear as a
characteristic property of quantum systems. The study of this issue should
serve us to establish contact with one of the distributions common to quantum
statistical theory.

\subsection{Including temperature-independent fluctuations}

The distribution appropiate for considering all fluctuations cannot be of the
form of Eq. (\ref{22a}), as has been already established. Moreover in order to
generalize Eq. (\ref{521c}) to include zero-point fluctuations we should look
for a distribution $W_{s}(E)$ that maximizes the entropy and that yields%

\begin{equation}
(\sigma_{E}^{2})_{s}=U^{2} \label{140}%
\end{equation}
for every temperature, that is we require $(\overline{E^{2}})_{s}=2U^{2}$ (the
subscript $s$ denotes averaging with respect to $W_{s}$ to distinguish from
the mean values calculated with $W_{g}$). The demand (\ref{140}) is immediate
by considering that the fluctuations come about from the interferences among a
huge amount of independent modes, and therefore the central limit theorem applies.

According to the maximum entropy formalism,$^{\text{\cite{mef}}}$ a
distribution satisfying these constraints is given by
\begin{equation}
W_{s}(E)=\frac{1}{U}e^{-E/U}. \label{142}%
\end{equation}
Obviously the selection $U=\beta^{-1}$ $(\mathcal{E}_{0}=0)$ results in the
usual canonical distribution Eq. (\ref{23}) and leads to the classical
expression, Eq. (\ref{521c}). But with the temperature-independent energy
$\mathcal{E}_{0}\neq0,$ $U$ is given by Planck's spectrum and the resulting
total fluctuations are (with $U_{T}$ given by Eq. (\ref{556}))
\begin{equation}
(\sigma_{E}^{2})_{s}=U^{2}=(U_{T}+\mathcal{E}_{0})^{2}=U_{T}^{2}%
+2\mathcal{E}_{0}U_{T}+\mathcal{E}_{0}^{2}. \label{144}%
\end{equation}
This shows that Eq. (\ref{521c}) can indeed be generalized to include the
temperature-independent energy and fluctuations in the case $q\neq0$. In the
description afforded by $W_{s}$, at zero temperature the energy does not have
a fixed value but is allowed instead to fluctuate with variance $\mathcal{E}%
_{0}^{2}$; this term represents the temperature-independent fluctuations. For
the thermal fluctuations we obtain from Eq. (\ref{144}) (omitting the
subscript $s$)
\begin{equation}
\left(  \sigma_{E}^{2}\right)  _{T}=\sigma_{E}^{2}-\mathcal{E}_{0}^{2}%
=U_{T}^{2}+2\mathcal{E}_{0}U_{T}, \label{146}%
\end{equation}
in agreement with Eq. (\ref{550}), since both $W_{g}$ and $W_{s}$ yield the
same \emph{thermal} averages.

Let us now decompose the total energy into two fluctuating parts,
\begin{equation}
E=E_{T}+E_{0}. \label{154}%
\end{equation}
where these terms stand for the (fluctuating) thermal and
temperature-independent energies, respectively. The total fluctuations are
then
\begin{subequations}
\begin{equation}
\sigma_{E}^{2}=\sigma_{E_{T}}^{2}+\sigma_{E_{0}}^{2}+2\Gamma(E_{T}%
,E_{0})=U_{T}^{2}+2\mathcal{E}_{0}U_{T}+\mathcal{E}_{0}^{2}, \label{156}%
\end{equation}
where $\Gamma(E_{T},E_{0})$ stands for the covariance of its arguments,
\begin{equation}
\Gamma(E_{T},E_{0})\equiv\overline{E_{T}E_{0}}-\overline{E_{T}}~\overline
{E_{0}}. \label{156a}%
\end{equation}
To write the second equality in (\ref{156}) we used Eq. (\ref{144}). Using Eq.
(\ref{146}) for $\sigma_{E_{T}}^{2}$ and putting $\sigma_{E_{0}}%
^{2}=\mathcal{E}_{0}^{2}$ leads to $\Gamma(E_{T},E_{0})=0,$ which shows that
the fluctuations of $E_{T}$ and $E_{0}$ are statistically independent, as was
to be expected due to the independence of their sources.

The fact that $\sigma_{E}^{2}$ differs from zero at null temperature confirms
that $W_{s}$ is not limited to a thermodynamic description but affords a
statistical one, which includes fluctuations beyond the thermal ones. This
being the case, it is clear that the corresponding entropy $S_{s},$ defined
using $W_{s}$ in the relation (\ref{1a}), will not coincide with the thermal
entropy in Eq. (\ref{14}) derived from $W_{g},$ since the former should
accomodate the new source of fluctuations. Such entropy is given by
\end{subequations}
\begin{equation}
S_{s}=-k\int W_{s}\ln W_{s}dE=k\ln U+k, \label{148}%
\end{equation}
from which it follows that%
\begin{equation}
\frac{\partial S_{s}}{\partial U}=\frac{k}{U}. \label{150}%
\end{equation}
Comparison with the (thermal) entropy we have been using throughout the
previous sections, which satisfies
\begin{equation}
\frac{\partial S}{\partial U}=\frac{1}{T}, \label{152}%
\end{equation}
leads us to conclude that both entropies coincide only when $\mathcal{E}%
_{0}=0$ (and consequently $U=kT$). The existence of a fluctuating
temperature-independent energy can be accommodated for by introducing a
"statistical" entropy $S_{s}$, as was done here, or equivalently a (quantum)
redefinition of the temperature, $\beta\longrightarrow1/U(\beta)$. A detailed
discussion of these matters can be found in Ref. \cite{entropy}.

\subsection{Quantum fluctuations and zero-point energy}

We have seen that the statistical description afforded by the distribution
$W_{s},$ Eq. (\ref{142}), is linked to unfreezable zero-point fluctuations. On
the other hand it is clear by now that the zero-point energy is crucial in
going from a classical description to a quantum one. It therefore makes sense
to investigate how these temperature-independent fluctuations manifest
themselves in some statistical properties of the quantum systems. Here we
limit our inquiry to a most immediate aspect.

Let us focus our attention on the quadratures $q,p$ of the oscillator, related
to its energy according to%
\begin{equation}
E=(p^{2}+m^{2}\omega^{2}q^{2})/2m. \label{158}%
\end{equation}
To this end we should go over from the energy distribution given by Eq.
(\ref{142}) to a distribution $W(p,q)$ defined in the phase space $(q,p)$. We
first note that $W(E)$ stands for a reduced probability density in the
action-angle variables space $(E,\theta)$ (we omit the subscript $s$)%
\begin{equation}
W(E)=\int\limits_{0}^{2\pi}W(E,\theta)d\theta, \label{160}%
\end{equation}
where
\begin{equation}
W(E,\theta)dEd\theta=W(p,q)dpdq. \label{162}%
\end{equation}
Since $W(E,\theta)$ does not depend on $\theta$ for the equilibrium state,
Eqs. (\ref{158}) and (\ref{162}) lead to$^{\text{\cite{note3}}}$
\begin{equation}
W(p,q)=\frac{\omega}{2\pi U}\exp(-\frac{p^{2}+m^{2}\omega^{2}q^{2}}{2mU}).
\label{164}%
\end{equation}
This distribution, which is known in quantum theory as the \textit{Wigner
function},$^{\text{\cite{Wigner}}}$ can be factorized as a product of two
normal distributions,%
\begin{equation}
W(p,q)=W_{p}(p)W_{q}(q)=\frac{1}{\sqrt{2\pi\sigma_{p}^{2}}}e^{-\frac{p^{2}%
}{2\sigma_{p}^{2}}}\times\frac{1}{\sqrt{2\pi\sigma_{q}^{2}}}e^{-\frac{q^{2}%
}{2\sigma_{q}^{2}}}, \label{166}%
\end{equation}
where we have identified the variances $\sigma_{p}^{2}=$ $mU$ and $\sigma
_{q}^{2}=U/m\omega^{2}$. Thus we get that
\begin{equation}
\sigma_{q}^{2}\sigma_{p}^{2}=\frac{U^{2}}{\omega^{2}}=\frac{\mathcal{E}%
_{0}^{2}}{\omega^{2}}+\frac{\sigma_{E_{T}}^{2}}{\omega^{2}}\geq\frac
{\mathcal{E}_{0}^{2}}{\omega^{2}}=\frac{\hbar^{2}}{4}, \label{168}%
\end{equation}
where we have used Eq. (\ref{521b}) (with $\sigma_{E}^{2}$ written
appropriately as $\sigma_{E_{T}}^{2}$) to write the second equality, and
$\mathcal{E}_{0}=\hbar\omega/2$ in the last one. We see that the magnitude of
$\sigma_{q}^{2}\sigma_{p}^{2}$ is bounded from below because of the
fluctuations of the zero-point energy (the minimum value $\hbar^{2}/4$ is
reached when all thermal fluctuations have been suppressed). Eq. (\ref{168})
allows to identify the origin of the Heisenberg inequalities with the presence
of a \emph{fluctuating} zero-point energy, and hence the descriptions afforded
by thermal distributions such as $W_{g}$ cannot account for their meaning,
least of its origin. This result stresses again the fact that once a
zero-point energy has been introduced into the theory, new distributions
(specifically statistical rather than thermodynamic) are needed in order to
include its fluctuations and to obtain the corresponding quantum statistical
properties. It is also important to note that according to the present
discussion, the Heisenberg inequalities should be understood as referring to
ensemble averages, due to the statistical nature of Eq. (\ref{168}) --- or
possibly to time-averaged quantities, if the system satisfies an ergodic principle.

\begin{acknowledgement}
One of the authors (AVH) acknowledges finantial support from the Consejo
Nacional de Ciencia y Tecnolog\'{\i}a under Grant No. 191914.
\end{acknowledgement}

\section{Appendix A. Alternate method for determining the parity of
$\sigma_{E}^{2}(U).$}

We have seen (Section \ref{spectrum}) that the invariance of $\sigma_{E}^{2}$
under the inversion $E\rightarrow-E$ determines the parity of the function
$\sigma_{E}^{2}(U),$ thus eliminating the parameter $a_{1}$ in Eq.
(\ref{400}). Here we present an alternate method to arrive at the same result,
which allows to uncover the origin of this symmetry in both the classical and
quantum cases.

We start from Eq. (\ref{400})
\begin{equation}
\sigma_{E}^{2}(U)=a_{0}+a_{1}U+a_{2}U^{2}. \label{A5}%
\end{equation}
Together with Eq. (\ref{37}) this gives%
\begin{equation}
\frac{dU}{a_{0}+a_{1}U+a_{2}U^{2}}=-d\beta. \label{A7}%
\end{equation}
By following a similar procedure to the one that lead from Eq. (\ref{430}) to
Eq. (\ref{445}) we find that (\ref{445}) generalizes to
\begin{equation}
U(\beta)=\left\{
\begin{array}
[c]{lc}%
\frac{1}{a_{2}\beta}-\frac{a_{1}}{2a_{2}}, & \text{for }q=0;\\
\frac{\sqrt{q}}{2a_{2}}\coth\frac{\sqrt{q}}{2}\beta-\frac{a_{1}}{2a_{2}}, &
\text{for }q>0.
\end{array}
\right.  q\equiv a_{1}^{2}-4a_{0}a_{2}. \label{A9}%
\end{equation}
The roots of the equation $\sigma_{E}^{2}=0$ are now%
\begin{equation}
U_{\pm}=\frac{-a_{1}}{2a_{2}}\pm\frac{\sqrt{q}}{2a_{2}}. \label{A10}%
\end{equation}
As we have seen $\mathcal{E}_{0}$ is one of the roots $U_{\pm}$ and $a_{2}>0,$
thus $\mathcal{E}_{0}$ corresponds to the largest of both roots (as before,
the other one being unphysical), that is
\begin{equation}
\mathcal{E}_{0}=\frac{-a_{1}}{2a_{2}}+\frac{\sqrt{q}}{2a_{2}}. \label{A11}%
\end{equation}
\qquad If the theory admits a null value for $\mathcal{E}_{0}$ the demand of
finite relative dispersion at every temperature compels us to set $a_{0}$ and
$a_{1}$ in (\ref{390}) equal to $0.$ In this case $q=0$, and from
(\ref{A9})\ we obtain $U=(a_{2}\beta)^{-1}.$ As was done in Section
\ref{spectrum} comparison with the classical result leads to equipartition,
Eq. (\ref{421}).

If instead the theory allows for a zero-point energy $\mathcal{E}_{0}\neq0$,
$q$ can in principle acquire a value different from $0$. Thus taking $q>0$
gives using Eqs. (\ref{A9}) and (\ref{A11})
\begin{align}
U(\beta)  &  =\left(  \mathcal{E}_{0}+\frac{a_{1}}{2a_{2}}\right)  \coth
a_{2}\left(  \mathcal{E}_{0}+\frac{a_{1}}{2a_{2}}\right)  \beta-\frac{a_{1}%
}{2a_{2}}\label{A15}\\
&  =\mathcal{E}_{0}-\frac{\sqrt{q}}{2a_{2}}+\frac{\sqrt{q}}{2a_{2}}\coth
\frac{\sqrt{q}}{2}\beta.\nonumber
\end{align}
Incidentally notice that this expression together with Eq. (\ref{37}) gives
the dispersion as a function of $\beta,$%

\begin{equation}
\sigma_{E}^{2}=-U^{\prime}=\frac{q}{4a_{2}}\left(  \coth^{2}\frac{\sqrt{q}}%
{2}\beta-1\right)  \text{.} \label{A17}%
\end{equation}
\qquad\bigskip

On the other hand, Wien's displacement law Eq.(\ref{12}), namely%
\begin{equation}
U(\omega,T)=\omega f(\omega/T), \label{12bis}%
\end{equation}
is equivalent to the relation%
\begin{equation}
\left(  \frac{\partial U}{\partial\omega}\right)  _{T}-\frac{U}{\omega}%
=-\frac{T}{\omega}\left(  \frac{\partial U}{\partial T}\right)  _{\omega}.
\label{A19}%
\end{equation}
Taking into account that $\mathcal{E}_{0}=$~const$\times\omega$ and
substituting (\ref{A15}) in the last equation we obtain%
\[
a_{2}\mathcal{E}_{0}(\mathcal{E}_{0}+\frac{a_{1}}{2a_{2}})\beta+\frac{a_{1}%
}{2a_{2}}\left[  \frac{1}{2}\sinh2a_{2}(\mathcal{E}_{0}+\frac{a_{1}}{2a_{2}%
})\beta-\sinh^{2}a_{2}(\mathcal{E}_{0}+\frac{a_{1}}{2a_{2}})\beta\right]
\]%
\begin{equation}
=a_{2}(\mathcal{E}_{0}+\frac{a_{1}}{2a_{2}})^{2}\beta. \label{A21}%
\end{equation}
Since we are assuming $q\neq0$ (and thus $\mathcal{E}_{0}+a_{1}/(2a_{2})\neq
0$) it follows that $a_{1}\ $must be $0$ for this equation to be satisfied
irrespective of the temperature. Further, Eq. (\ref{A15}) reduces to Eq.
(\ref{508}), and therefore, taking the high-temperature limit to set $a_{2}=1$
as before, to Planck's law.

It is interesting to observe that for $\mathcal{E}_{0}>0,$ the even parity of
the dispersion $\sigma_{E}^{2}$ as a function of $U$ is a direct consequence
of Wien's law (which entails $\mathcal{E}_{0}\propto\omega)$. In the classical
case ($\mathcal{E}_{0}=0$) it was the demand that the relative dispersion
remains finite at $T=0$ which lead us to a similar conclusion about the parity
of $\sigma_{E}^{2}.$

\section{Appendix B. The value of $\overline{\text{ln }g(E)}.$}

We have seen that the mean value of a general function $f(E)$ calculated with
$W_{g}$ is%
\begin{equation}
\overline{f(E)}=\frac{1}{Z_{g}}\sum_{m}f(E_{m})e^{-\beta E_{m}}=\sum
_{m}f(E_{m})w_{m}. \label{B10}%
\end{equation}
Let us apply this to the function $g$ given by Eq. (\ref{136b}),%
\begin{equation}
f(E)=\ln g(E)=\ln\sum_{n=0}^{\infty}2\mathcal{E}_{0}\delta(E-E_{n}),
\label{b11}%
\end{equation}
whence%
\begin{equation}
\overline{\ln g(E)}=\sum_{m}\ln\left[  \sum_{n=0}^{\infty}2\mathcal{E}%
_{0}\delta(E_{m}-E_{n})\right]  w_{m}. \label{b12}%
\end{equation}
We write the single contribution for $n=m$ as $1/\varepsilon$ with
$\varepsilon$ $\rightarrow0.$ Since $\varepsilon$ is a numeric constant
independent of $m,$ it can be taken out of the sum, which results in%
\begin{equation}
\overline{\ln g(E)}=\ln\frac{1}{\varepsilon}\sum_{m}w_{m}=\ln\frac
{1}{\varepsilon}. \label{b13}%
\end{equation}
We verify that $\overline{\ln g(E)}$ is a numeric constant independent of
$\beta,$ as asserted in the main text.

That $\overline{\ln g(E)}$ should be a constant follows also from the
observation that \ according to the formalism of maximum
entropy$^{\text{\cite{mef}}}$ the distribution
\begin{equation}
W_{g}=\frac{1}{Z}e^{-\beta E+\ln g(E)}\label{b14}%
\end{equation}
corresponds to a simultaneous stationary value of the entropy and of $\ln
g(E).$


\begin{thebibliography}{99}
\bibitem {unity}A comment may be in place here. As demonstrated\ in Ref.
\cite{relat} the single dependence of the zero-point energy on frequency
allowed by special relativity is $\mathcal{E}_{0}(\omega)\sim\omega$. The
above classical thermodynamic calculation leads to the same expression, Eq.
(\ref{18}). In its turn the Schr\"{o}dinger equation of non relativistic
quantum mechanics gives for the ground state of a material harmonic oscillator
a similar prediction. We find here a nice example of the unity of physics ---
or rather, of nature --- strongly suggesting that the different chapters of
physics are but branches of a general theory to be.
\bibitem {note1}Indeed Einstein considered the distribution (\ref{22a})
assuming from the start the form of the function $g(E)$ (as dictated by the
quantization discovered by Planck). Here we will determine it as a result of
the theory, moving just in the opposite sense.
\end{thebibliography}
\end{document}